\documentclass[preprint,letterpaper,sort&compress,12pt]{elsarticle}

\makeatletter
\def\ps@pprintTitle{%
  \let\@oddhead\@empty
  \let\@evenhead\@empty
  \let\@oddfoot\@empty
  \let\@evenfoot\@oddfoot
}
\makeatother
 
\usepackage{hyperref}
\hypersetup{
    colorlinks,
    citecolor=blue,
    filecolor=blue,
	    linkcolor=blue,
    urlcolor=blue
}

\newcommand {\beg}{\begin{equation}}
\newcommand {\en}{\end{equation}}
\newcommand {\lb}[1]{\label{#1}}
\newcommand {\begar}{\begin{array}}
\newcommand {\enar}{\end{array}}
\newcommand {\integ}{\displaystyle{\int}}
\newcommand {\dis}{\displaystyle}

\begin{document}

 \title{Multi-soliton solutions of the two-dimensional matrix Davey--Stewartson equation}

\vspace{1cm}

\author{Andrei N. Leznov}
\address{Institute for High Energy Physics,142284 Protvino,
Moscow Region, Russia}

 \author{Emil A. Yuzbashyan\footnote{Email: eyuzbash@physics.rutgers.edu}}
 \address{Bogoliubov Laboratory of Theoretical Physics, Joint Institute for
Nuclear Research,
141980 Dubna, Moscow Region, Russia}

\begin{abstract}
 We explicitly obtain the 
$m$-soliton solutions of the (1+2)-dimensional
matrix Davey-Stewartson equation from the
known general solution of the matrix Toda chain with  fixed ends.
We write these solutions  in terms of $m+m$ independent solutions of a pair of linear
Shr\"odinger equations with Hermitian potentials.
\end{abstract}


\maketitle 

\section{Davey-Stewartson equation}
\label{intro}

 Let $u$ and $v$ be two non--singular $s\times s$ matrix functions of
$x,$ $y$, and $t$, i.e. each matrix element is a function of
 the $x$ and $y$ coordinates of the
two-dimensional space and the time $t$. Partial derivatives of these functions up to a
sufficiently  high order are assumed to exist.

 We define the matrix Davey--Stewartson equation  as the following
partial differential equation:
\beg
iu_t+au_{xx}+bu_{yy}-2au\int dy(u^\dagger u)_x-2b\int dx (u u^\dagger)_y \cdot u=0,
\lb{dse}
\en
where $a$ and $b$ are arbitrary real numbers and
$z^\dagger$ stands for the Hermitian conjugate of a matrix $z$.
It will be convenient to deal not with  Eq.~(\ref{dse})
but with the following expanded system, which we  call the matrix
Davey--Stewartson system:
\beg
\begar{r}
iu_t+a u_{xx}+b u_{yy}-2au\integ dy (vu)_x-2b\integ dx (uv)_y\cdot u=0,\\\\
-iv_t+a v_{xx}+b v_{yy}-2a\integ dy (vu)_x\cdot v-2b v\integ
dx (uv)_y =0.
\enar\lb{dss}
\en
Bellow, for definiteness we set  $a=b=1$. It is easy to see that the
matrix Davey--Stewartson equation is the
system~(\ref{dss}) under the additional condition 
\begin{equation}
v=u^\dagger,
\label{reality}
\end{equation}  
which we call the reality condition.

In the case $s=1$~(scalar case), when $u$ and $v$ are scalar functions and the
order of the multipliers is not essential, Eq.~(\ref{dse})
is the usual, well--known Davey--Stewartson equation \cite{2}, for which
the soliton solutions   were obtained in \cite{2a}.

\section{Discrete substitution}

We use the discrete symmetry (transformation, substitution) approach~\cite{3, 3_1} to solve the problem. In the present case, the relevant discrete symmetry   is the following change of the unknown matrices $u$  and $v$:
\beg 
\tilde u=v^{-1},\quad \tilde v=\left[vu-(v_x v^{-1})_y\right]
v\equiv v \left[uv-(v^{-1}v_y)_x\right].
\lb{ds} 
\en
Here  $\tilde u$ and $\tilde v$ denote  the ``new'', transformed matrices. By direct calculation it can be checked that Eq.~(\ref{dss}) is invariant
with respect to this transformation, i.e. matrices $\tilde u$ and $\tilde v$ satisfy the same
system~(\ref{dss}) as the  matrices $u$ and $v$ do. 

Mapping~(\ref{ds}) is 
invertible and the ``old'' matrices $u$ and $v$ can be expressed trough the
``new'' ones as
\beg
v=\tilde u^{-1},
\quad u=\left[\tilde u\tilde v-(\tilde u_y \tilde u^{-1})_x\right] \tilde u
\equiv \tilde u \left[\tilde v\tilde u-(\tilde u^{-1}\tilde u_x)_y\right].
\lb{is}
\en
Further, the transformation~(\ref{ds})  gives rise to two equivalent recurrence relations
\beg
\left((v_n)_x v_n^{-1}\right)_y=v_n v_{n-1}^{-1}-v_{n+1} v_n^{-1},\quad
u_{n+1}= v_n^{-1},
\lb{ch1}
\en
and
\beg \left(v_n^{-1} (v_n)_y\right)_x=v_{n-1}^{-1}v_n-v_n^{-1}v_{n+1},
\lb{ch2}
\en
where $(v_n, u_n)$ is the
result of  $n$ applications of the transformation~(\ref{ds})  to a   pair of initial matrices
$(v_0, u_0)$.  Recurrence relations~(\ref{ch1}) and (\ref{ch2}) with  boundary
conditions $v_{-1}^{-1}=v_N=0$ define the matrix Toda chain with  fixed ends~\cite{1}.

We note that the (1+1)-dimensional version of the mapping~(\ref{ds}) is mentioned in \cite{5}.
In the scalar case $s=1$,  the general solution of the Toda chain with
 fixed ends was found in~\cite{6} for all  series of semi-simple algebras,
except $E_7$ and $E_8$. In \cite{7} this result was reproduced in terms
of the invariant root technique applicable to all semi-simple series.

The explicit general solution of the  matrix Toda chain with  fixed ends
has been obtained in~\cite{1}. The solution provides $v_0$ such that $v_{-1}^{-1}=v_N=0$ in terms of $N+N$ arbitrary independent
matrix functions of a single argument~$X_r(x),Y_r(y)$,
\beg
v_0=\sum_{r=1}^{N} X_r Y_r. 
\lb{dT}
\en
The $u_N$ that corresponds to this $v_0$ has the form
\beg
u_{N}=\sum_{r=1}^{N} \tilde Y_r(x)\tilde X_r(y). 
\lb{iT}
\en
Here the matrices  $\tilde X$ and $ \tilde Y$  are  not arbitrary, but 
depend on  $ X$ and  $Y$ in a certain way. Both these results (\ref{dT}) and (\ref{iT})
will be used below.

\section{General strategy}
\label{strategy}

Our goal is to solve the Davey--Stewartson equation~(\ref{dse}), or, equivalently, the  Davey--Stewartson system~(\ref{dss}) under the condition of reality $u=v^\dagger$.
Here, we describe how the discrete transformation is used for that.
The general idea is as follows. Start with an obvious solution of the
Davey-Stewartson system. This typically does not satisfy the reality condition and, therefore, is not a solution of the Davey-Stewartson equation. However, by repeatedly applying the discrete transformation~(\ref{ds}) to this initial, obvious solution, we generate a solution which does satisfy this condition and, therefore, is a solution of the Davey-Stewartson equation.

For $u_0=0$ the first equation of the system (\ref{dss}) is satisfied
identically and the second one gives
\beg
-i{v_0}_t+ {v_0}_{xx}+{v_0}_{yy}+V_1(t,x)
v_0+v_0 V_2(t,y)=0,
\lb{reduced}
\en
where $V_1$ and $V_2$ are arbitrary
$s\times s$ matrix functions of their arguments [these terms arise from
the indefinite integrals $\int dx (uv)_y$, $\int dy (uv)_x$ in the
system~(\ref{dss})]. Obviously, the condition of reality does not hold for
this solution. But
after applying the  transformation~(\ref{ds}) a sufficient  number of times, 
it is possible to arrive at a solution  which satisfies it.

To clarify this,  consider a solution $(u,v)$ for which
the condition of reality holds, i.e. $u=v^\dagger$. Let
$(u_1,v_1)$ and $(u_{-1},v_{-1})$ be the results of the direct~(\ref{ds}) and
inverse~(\ref{is}) transformations, respectively. It is straightforward to show that
$u_{-1}={v_1}^\dagger$ and $v_{-1}=u_1^\dagger$.  Similarly, after $m$ steps, we have
$u_{-m}={v_m}^\dagger$ and $v_{-m}=u_m^\dagger$, where the index $m$ $(-m)$
stands for the result
of $m$ direct (inverse) transformations. And vice versa, one can prove
that if we start with a solution $(u_0=0, v_0)$ and after
$2m$ direct discrete transformation obtain $(u_{2m}=v^\dagger_0, v_{2m}=0)$, the
solution in the middle of the chain automatically satisfies the reality
condition, i.e. $u_{m+1}=v_{m+1}^\dagger$.

 Equations $u_0=v_{2m}=0$ are already solved by
 formula (\ref{dT}). Therefore, it remains to solve the equation $u_{2m}=v^\dagger_0$.
This equation leads to the following relations between $X_r$ and  $\tilde X_{\sigma[r]}$ and
between $Y_r$ and $\tilde Y_{\sigma[r]}$:
\beg
X_r^*=\tilde X_{\sigma[r]}\qquad Y_r^*=\tilde Y_{\sigma[r]},
\lb{cr}
\en
where $\sigma$ denotes one of the $(2m)!$ possible permutations of the
$2m$ lower
indices. To solve~(\ref{cr}), we first  need to determine the dependence
of $\tilde X$ and $\tilde Y$ on $X$ and $Y$, respectively.
Finally, Eq.~(\ref{reduced}) in terms of $X_r$ and $Y_r$ reads
\beg
-i{X_r}_t+ {X_r}_{xx}+V_1(t,x) X_r=0,\quad -i{Y_r}_t+ {Y_r}_{xx}+Y_r V_2(t,y)=0.
\lb{reducedtoX}
\en
Thus,  constructing  $m$-soliton solutions of  the Davey--Stewartson equation~(\ref{dse}) involves
  the following steps:
\begin{itemize}
\item determine the functions  $\tilde X_i\left(X_1,\dots,X_{2m}\right)$ and
$\tilde Y_i\left(Y_1,\dots Y_{2m}\right)$;
\item solve the system~(\ref{cr});
\item determine the time dependence of matrix functions $X_r$ and $Y_r$  that solves the system of equations~(\ref{reducedtoX}).
\end{itemize}
After this, substituting $X_r$ and $Y_r$ into~(\ref{dT}), we obtain $v_0$, such that
$u_{m+1}=v_{m+1}^\dagger$ is a particular ($m$-soliton) solution of the
Davey--Stewartson equation~(\ref{dse}).

\section{Scalar case}
\label{scalar_case}

To gain some experience, first we  consider the
scalar case $s=1$ for which most of the necessary steps   are
 known and much simpler then in the general matrix case.

In this case, for the aforementioned  boundary conditions the following
formulas hold for arbitrary $k$ \cite{4, 4_1}:
\beg
u_k={\mathrm{Det}_{k-1}\over \mathrm{Det}_k},\quad v_k={\mathrm{Det}_{k+1}\over \mathrm{Det}_k},
\quad \mathrm{Det}_{-1}\equiv0,\quad \mathrm{Det}_0\equiv1,
\lb{Detk}
\en
where $\mathrm{Det}_k$ is  the principle minor of dimension $k$
 of the matrix ($v^0\equiv v_0$)
$$
\pmatrix{    v^0 & v^0_x    & v^0_{xx}  &....\cr
           v^0_y & v^0_{xy} & v^0_{xxy} &.....\cr
        v^0_{yy} & v^0_{xyy} & v^0_{xxyy} &.....\cr
              ...& ..........&  ..........&.....\cr
              ...& ..........&  ..........&.....\cr},
$$
and $v^0$ is determined by~(\ref{dT}), where $X_r$ and $Y_r$ are
arbitrary scalar
functions of their arguments. Substituting~(\ref{dT}) into the expression for
$u_{2m}$ from~(\ref{Detk}) and comparing with~(\ref{iT}), we find
\beg
\tilde X_r(x)={W_{2m-1}( X_1,X_2,...,X_{r-1},X_{r+1},...X_{2m})\over
W_{2m}(X_1,X_2,....X_{2m})}.
\lb{scX}
\en
Here and below  $W_k$ is the Wronskian,  
\beg
W_k(g_1,\dots,g_k)\equiv
\begar{|llll|}
g_1&g_2&\ldots&g_k\\g_1'&g_2'&\ldots&g_k'\\\vdots&\vdots&\ddots&\vdots\\
g_1^{(k-1)}&g_2^{(k-1)}&\ldots&g_k^{(k-1)}\end{array}\,\, ,\qquad W_0\equiv1,
\lb{wr}
\en
and the prime indicates the derivative with respect to the spatial coordinate on which the function depends ($x$ in this case).
Expressions for $\tilde Y_r$ obtain from~(\ref{scX}) by simple
exchange $X\to Y$ and $x\to y$.

In the condition of reality~(\ref{cr}) we use the permutation
$\sigma[r]=2m-r+1$.
To resolve~(\ref{cr}) and~(\ref{reducedtoX}), it is convenient to represent
the functions $X_r$ and $Y_r$ in the Frobenious-like form
\beg
\begar{l}
X_1=\phi_1,\quad X_r=\phi_1 \integ dx\, \phi_2\dots\integ dx\, \phi_r,\\\\
Y_1=\psi_1,\quad Y_r=\psi_1 \integ dy\, \psi_2\dots\integ dy\, \psi_r .\\
\enar
\lb{frob}
\en
Eq.~(\ref{scX}) implies
\beg
\tilde X_{2m}=\Bigl(\prod^{2m}_{k=1}\phi_k\Bigr)^{-1},\quad \tilde X_r=
\Bigl(\prod^{2m}_{k=1}\phi_k\Bigr)^{-1}\int
dx\,\phi_{2m} \dots\int dx\,\phi_{2m-r}.
\lb{tilde}
\en
Now the reality condition~(\ref{cr}) takes the  form
\beg
\phi^*_r=\phi_{2m-r+2}\quad (r=2,3,...2m), \quad
\phi_{m+1}=\phi^*_{m+1}=\Bigl(\prod_{k=1}^m \phi_k
\phi^*_k\Bigr)^{-1},
\lb{crphi}
\en
where the asterisk denotes the complex conjugate.
From~(\ref{reducedtoX}) we have
\beg
{\phi_r}_t=\biggl(\phi_r \Bigl(\ln\phi_r \prod_{k=1}^{r-1} \phi_k
^2\Bigr)'\biggr)'.
\lb{reducedtophi}
\en
 The imaginary unity~$i$ here has been absorbed into the time
variable, which, therefore, should be treated as  a pure imaginary
from now on. It is straightforward  to    check that  
Eqs.~(\ref{crphi}) and (\ref{reducedtophi})
are compatible. In particular, if Eq.~(\ref{reducedtophi})  holds for   $\phi_r$ with any $r\le m$,  it also holds for $\phi_{2m-r+2}$. Hence, it is sufficient to
consider only  $r\le m$ in Eq.~(\ref{reducedtophi}).

Now let us
  introduce new unknown functions $f_r^{-1}=\phi_1\cdots\phi_r,
\quad r\le m+1$.  From~(\ref{reducedtophi}) we find
\beg
({f_r^{-1}f_{r-1}})_t=-\left({f_r^{-1}f_{r-1}} (\ln f_r f_{r-1})'\right)'.
\lb{reducedtof}
\en
Further, Eq.~(\ref{crphi}) implies
$f^*_m=f_{m+1}^{-1}$. Substituting this into  
 Eq.~(\ref{reducedtof}) for $r=m+1$, we obtain
\beg
(f_m f^*_m)_t=\left(f_m f^*_m\left(\ln f_m {f^*_m}_{-1}\right)'\right)'.
\lb{m-th}
\en
Eq.~(\ref{m-th}) is equivalent to the one-dimensional Shr\"odinger
equation with an arbitrary real potential $U$,
\beg
{f_m}_t+f_m''=Uf_m,\quad U=U^*.
\label{shr}
\en
Next,  set $r=m$ in Eq.~(\ref{reducedtof}),
\beg
({f_m^{-1}f_{m-1}})_t=-\left({f_m^{-1}f_{m-1}} (\ln f_m f_{m-1})'\right)'.
\lb{m-1-th}
\en
This equation implies
\beg
f_m^{-1}f_{m-1}=z',\quad (\ln f_m f_{m-1})'=-{z_t\over z'},
\en
where $z$ is a   function of $x$ and $t$.
Eliminating   $f_{m-1}$, we observe that the function $zf_m$ satisfies
exactly the same Eq.~(\ref{shr}) as $f_m$ does.
Let $u_i$  $(1\le i\le m)$ be $m$ independent solutions of~(\ref{shr}). Then,
\beg
f_m=u_1,\quad zf_m=u_2\Longrightarrow  f_{m-1}=z'f_m=
{\begar{|ll|}u_1&u_2\\u_1'&u_2'\enar\over u_1}.
\lb{1step}
\en
 In the general case, for arbitrary $i$ the following formula holds:
\beg
f_r={W_{m-r+1}\over W_{m-r}},\quad r\le m,
\lb{generalstep}
\en
where $W_i=W_i(u_1,\dots,u_i)$.

To prove~(\ref{generalstep}), we use the well-known Jacobi identity for
determinants. Consider a matrix $T$    infinite in both directions. Let
$D_n(T)$ be the determinant of its
$n\times n$ principle minor and let $T^s$ and $T_p$ be the matrices obtained from
$T$ by deleting its $s$th column and   $p$th row, respectively.
In these
notations the Jacobi identity takes the form
\beg
D_n(T)D_n(T^n_n)-D_n(T^n)D_n(T_n)=D_{n+1}(T)D_{n-1}(T).
\label{jacobi}
\en
This identity in turn implies
\beg
W_i\overline{W}_i'-W_i'\overline{W}_i=W_{i-1}W_{i+1},
\label{jacobiderived}
\en
where $\overline{W}_i=W_i(u_i\to u_{i+1})=W_i(u_1,u_2,\dots,u_{i-1},u_{i+1})$.

 Eq.~(\ref{reducedtof}) for arbitrary $r$ implies
\beg
f_{r-1}=z'f_r, \quad (\ln f_rf_{r-1})'=-{z_t\over z'}.
\lb{part}
\en
Similarly to the $r=m$ case,  eliminating $f_{r-1}$, we find that $f_r$ and $zf_r$ are
two different solutions of the same equation. And if $f_r$ is given by Eq.~(\ref{generalstep}),  then
$$
zf_r={\overline{W}_{m-r+1}\over W_{m-r}}.
$$
Now  using
the identity~(\ref{jacobiderived})  and Eq.~(\ref{part}), we find   
$$
f_{r-1}={W_{m-r+2}\over W_{m-r+1}}.
$$
Thus, we have proved formula~(\ref{generalstep})   by induction.

Finally, for functions $\phi_r$ from the definition of $f_r$ and
Eqs.~(\ref{crphi}), we have
\beg
\begar{ccc}
\phi_{m+1}=v_1v^*_1,&\phi_r=\dis{{W_{m-r+2} W_{m-r}\over W_{m-r+1}^2}},&\\&&\\
\phi_1=\dis{{W_{m-1}\over W_m}},&\phi^*_r=\phi_{2m-r+2},&r\le m.
\enar\lb{phi}
\en

Analogues expressions   for functions $\psi_k$ are
\beg
\begar{ccc}
\psi_{m+1}=v_1v^*_1,&\psi_r=\dis{{W_{m-r+2} W_{m-r}\over W_{m-r+1}^2}},&\\&&\\
\psi_1=\dis{{W_{m-1}\over W_m}},&\psi^*_r=\psi_{2m-r+2},&r\le m.
\enar\lb{psi}
\en
In~(\ref{psi}), $W_i=W_i(v_1,\dots,v_i)$ and
$v_i\equiv v_i(y),$ $1\le i\le m,$
are $m$ independent solutions of the $(1+1)$--dimen\-si\-onal
linear Shr\"o\-din\-ger equa\-tion with an arbit\-rary real potential $V$,
\beg
{v_i}_t+v_i''=Vv_i,\quad V=V^*.
\lb{shyy}
\en

\section{Matrix case}
\label{matrix_case}

In this section, we solve the general problem as it has been formulated in  Sections~\ref{intro}
and \ref{strategy}, i.e. we construct $m$-soliton solutions of the matrix  Davey-Stewartson equation for an arbitrary
dimension of the
unknown matrix $u$.  We thus obtain matrix generalizations of the results of the
previous section.  In particular, it can be seen  that regular determinants are replaced by quasi-determinants of matrices with
non-commutative entries, a notion  introduced recently by
Gelfand and Retarh~\cite{quasi,quasi_1}.
Even though we will use a different technique which is more appropriate in our particular case,
 quasi-determinants can be used as well.

With the chain~(\ref{ch1},\ref{ch2}) under the above-mentioned  boundary conditions
we  associate  the following recurrence relations:
\beg
R_n\equiv v_n^{-1} {v_n}_y,\quad S_n^q\equiv \sum_{k=0}^{n-1}
 \left(S^{q-1}_{k_y}+R_k S^{q-1}_k\right),
 \lb{recdef}
\en
with the boundary conditions $S^0_i\equiv 1$ for all $i$.
It follows from the definitions~(\ref{recdef}) and Eqs.~(\ref{ch1}) and (\ref{ch2}) that
\beg
S^1_n=\sum_{k=0}^n R_n,\quad S_0^q=
v_0^{-1}{v_0}_{\underbrace{y\cdots y}_{\scriptstyle q}},
\lb{unsderb}
\en
\beg
v_{n+1}=-v_n (S_{n+1}^1)_x=(-1)^{n+1}v_0 (S_1^1)_x(S_2^1)_x\cdots(S_{n+1}^1)_x,
\lb{mdetk}
\en
and   
\beg
S_n^q=\left[(S_{n-1}^1)_x\right]^{-1} ( S_{n-1}^{q+1})_x.
\lb{mainrec}
\en

Now let us find the dependence of $\tilde X$ on $X$. For this, we use the
fact that
each matrix function $X_i$ is determined only by matrices $X_1,\dots,X_{2m}$
and therefore we can choose matrices $Y_1,\dots,Y_{2m}$  arbitrarily.
It is convenient to choose
\beg
Y_i={y^{i-1}\over (i-1)!}E, \quad v_0=X_1+{y\over1!}X_2+\cdots+{y^{2m-1}\over
(2m-1)!}X_{2m},
\lb{choose}
\en
where $E$ is the $s\times s$ identity matrix. Substituting~(\ref{choose}) into the
expression for $v_{2m-1}$ from~(\ref{mdetk}), we find
\beg
\left.v^{-1}_{2m-1}\right|_{y=0}=
-\left[X_1(T^1_0)_x(T^1_1)_x\cdots(T^1_{2m-2})_x\right]^{-1},
\lb{v2m}
\en
where
\beg
T_n^q=\left[(T_{n-1}^1)_x\right]^{-1}(T_{n-1}^{q+1})_x,
\lb{recT}
\en
and the boundary conditions read
\beg
T^q_0=S_0^q|_{y=0}=X_1^{-1}X_{q+1}.
\en
Expression~(\ref{v2m}) corresponds to one of the functions $\tilde X_i$.  Without loss of generality, we can choose this to be $\tilde X_1$, i.e.,
\beg
(\tilde X_1)^{-1}=X_1 (T^1_0)_x(T^1_1)_x\cdots(T^1_{2m-2})_x\equiv
F(x_1,\dots,x_{2m}).
\lb{X}
\en
The answer for an arbitrary $i$ can be derived from~(\ref{X}) by a cyclic
permutation of the indices,
\beg
\tilde X_i^{-1}=(-1)^{i-1}F(\sigma_i[x_1,\dots,x_{2m}]).
\lb{cyrcleX}
\en
An arbitrary overall multiplier can be added to the right hand side of Eq.~(\ref{cyrcleX}), which will correspondingly modify
the expression for $\tilde Y_i$. We added $(-1)^{i-1}$ for
future  convenience. 

Further, using (\ref{cyrcleX}) and (\ref{mdetk}), we obtain
\beg
\tilde Y_1^{-1}=-(Q^1_{2m-2})_y\cdots(Q^1_0)_y,\quad
Y_1\equiv G(Y_1,\ldots,Y_{2m}),
\lb{Y}
\en
\beg
\tilde Y_i^{-1}=(-1)^iG(\sigma_i[Y_1,\ldots,Y_{2m}]),
\lb{cycler}
\en
where
\beg
Q^s_n= (Q^{s+1}_{n-1})_y\left[(Q^1_{n-1})_y\right]^{-1},
\en
and
\beg
Q^s_0=Y_{s+1} Y_1^{-1}.
\en
As in the previous section, we write the initial functions $X_r$ and $Y_r$
in the Frobenious--like form,
\beg
\begar{lll}
X_1=\phi_1,&X_2=\phi_1\integ dx\phi_2,&X_3=\phi_1 \integ dx \phi_2\integ
dx\phi_3,\quad\ldots,\\&&\\
Y_1=\psi_1,&Y_2=\integ dx\psi_2\cdot\psi_1,&
Y_3=\integ dx\left(\integ dx\psi_3\cdot\psi_2\right)\cdot\psi_1,\quad\ldots
\enar
\lb{mfrob}
\en
 The corresponding expression for $\tilde X_r$
coincides with (\ref{tilde}), up to a permutation of indices. The only difference is that in the
matrix case
the order of the
multipliers must be taken into account. Specifically, we have
\beg
\begar{lll}
\tilde X_1=p,&\tilde X_2=\integ dx\phi_{2m}\cdot p,&
\tilde X_3=\integ dx\left(\integ dx\phi_{2m-1}\cdot\phi_{2m}\right)\cdot p,
\quad\ldots,\\&&\\
\tilde Y_1=-s,&\tilde Y_2=-s\integ dx\psi_{2m},&\tilde Y_3=-s\integ dx\psi_{2m}
\integ dx\psi_{2m-1},\quad\ldots,
\enar
\lb{mtilde}
\en
where $p=(\phi_1\cdots\phi_{2m})^{-1}$ and $s=(\psi_{2m}\cdots\phi_{1})^{-1}$.
The condition of reality taken in the form
$\tilde X_r=X_r^\dagger,\quad\tilde Y_r=Y_r^\dagger$ gives
\beg
\begar{l}
\phi^\dagger_r=\phi_{2m-r+2},\quad 2\le r\le m,\\\\
\phi^{-1}_{m+1}=(\phi^\dagger_{m+1})^{-1}= (\phi_1 \phi_2\cdots\phi_m)^\dagger
(\phi_1 \phi_2\cdots\phi_m),\\\\
\psi^\dagger_r=\psi_{2m-r+2},\quad 2\le r\le m,\\\\
\psi^{-1}_{m+1}=(\psi^\dagger_{m+1})^{-1}=-(\psi_m \psi_{m-1}\cdots\psi_1)^\dagger
(\psi_m \psi_{m-1}\dots\psi_1).\\
\enar
\lb{crpsi}
\en
The fact that all functions $X_r$ are solutions of the same
equation~(\ref{reducedtoX}) leads to the following system:
\beg
-(\phi_s)_t+(2(\phi_1 \phi_2 ...\phi_{s-1})^{-1} (\phi_1 \phi_2 ...\phi_{s-1})'
\phi_s+\phi_s')'=0.
 \lb{reducedtomphi}
\en
Introducing new 
functions $f_r^{-1}=\phi_1\phi_2\dots\phi_r$, 
we find from~(\ref{reducedtomphi})
\beg
-(f_{r-1} f_r^{-1})_t=\left[f_{r-1}'f_r^{-1}-f_{r-1}(f_r^{-1})'\right]'.
\label{reducedtomf}
\en
It follows from~(\ref{crpsi}) and (\ref{reducedtomf})   that $f_m$ is a
solution of the $(1+1)$-dimensional linear matrix Shr\"odinger equation with an arbitrary Hermitian
potential $W$,
\beg
{f_m}_t+f_m''=Wf_m,\quad W=W^\dagger.
\label{mshx}
\en
We solve the system of equations~(\ref{reducedtomf}) similarly to how we solved the system~(\ref{reducedtof}) in the
previous section, except that  now we use the recurrence relations  instead of the Jacobi identity. We have
\beg
f_1=u_1,\quad f_{m-r}=(U^1_r)'\cdots(U^1_0)'u_1.
\label{f}
\en
Matrix functions $U^q_n$ are determined by the recurrence relations
$$
U^q_n= (U^{q+1}_{n-1})'\left[(U^1_{n-1})'\right]^{-1},
$$
combined with the boundary conditions
$$
U^r_0=u_{r+1} u_1^{-1},
$$
where matrices $u_r$ are distinct solutions of  Eq.~(\ref{mshx}).

  Finally, for $\phi_{m-r}$ we
derive
\beg
\begar{lll}
\phi_1=f_1^{-1},&\phi_{m-r}=(U^1_r)',&0\le r\le m-2,\\&&\\
\phi_{m+1}=u_1u_1^\dagger,&\phi^\dagger_r=\phi_{2m-r+2},&2\le r\le m,
\enar
\lb{mphi}
\en
and for $\psi_i$, we have
\beg
\begar{l}
\psi_1^{-1}=v_1(V^1_{m-2})'\cdots(V^1_0)'v_1,\\\\
\psi_{m-r}=(U^1_r)',\quad0\le r\le m-2,\\\\
\psi_{m+1}=-v_1v_1^\dagger,\quad\psi^\dagger_r=\psi_{2m-r+2},\quad2\le r\le m,
\enar
\lb{mpsi}
\en
where
$$
V^q_n=\left[(V^1_{n-1})'\right]^{-1}(V^{q+1}_{n-1})',
$$
and
$$
 V^r_0=v_1^{-1}v_{r+1}.
$$
Matrices $v_i(y)$ are  distinct solutions of the
$(1+1)$-dimensional linear
Shr\"odinger equation with  an arbitrary Hermitian potential $M$.
$$
{v_i}_t+v_i''=v_iM,\quad M=M^\dagger.
$$
Substituting (\ref{mphi}) and (\ref{mpsi})  into (\ref{mfrob}) and
(\ref{dT}) and then into the formula for $v_{m+1}$ from~(\ref{mdetk}), we find
the $m$--soliton solution of the matrix Davey-Stewartson equation. We do not write the
explicit 
expression for the solution, because it can easily be derived, but is too long to be
presented here.

\section{Simplest example: one-soliton  solution}

Substituting $m=1$ into the appropriate formulas of the last section, we find
$$
v_0=X_1Y_1+X_2Y_2,\quad X_1=\phi_1,\quad X_2=\phi_1\int dx\phi_2,
$$
$$
Y_1=\psi_1,\quad Y_2=\int dx\psi_2\cdot\psi_1.
$$
This  produces the following expression for the one-soliton solution
of the matrix Davey-Stewartson equation:
\beg
u_1=\psi_1^{-1}\left(1+\int dx \phi_2\int dy \psi_2\right)^{-1}\phi_1^{-1}.
\label{ex}
\en
The matrix functions $\phi(t,x)$ and $\psi(t,y)$ are determined by the solutions $u$ and $v$, respectively, of
the one-dimensional linear Schr\"o\-din\-ger equations as follows:
$$
\begar{l}
\dis
\phi_1=u^{-1},\quad \phi_2=uu^\dagger,\\\\
\psi_1=v^{-1},\quad \psi_2=-v^\dagger v,\\\\
u_t+u_{xx}+uM_1(t,x)=0,\\\\
v_t+v_{yy}+M_2(t,y)v=0,\quad M_{1,2}=M^\dagger_{1,2}.\\\\
\enar
$$

\section{Conclusion}

The main result of the present paper is the explicit construction of the $m$-soliton
solutions of the (1+2)-dimensional matrix
Davey-Stewartson equation. These solutions are given in terms of $m+m$ independent solutions of a pair of linear
(1+1)-dimensional Shr\"odinger equations (see Sections~\ref{scalar_case} and \ref{matrix_case}).
From the group-theoretical point of view this means that we have realized
a finite-dimensional representation of the group of integrable mappings.
This viewpoint, however, remained outside of our concrete calculations.

Note that the restriction with the finite-dimensional matrices is  
nonessential. We have never used this restriction and, moreover, the size of the matrices
($s$)  did not appear in any of our expressions.


\begin{thebibliography}{**}


\bibitem{1}

A. N. Leznov and E. A. Yuzbashyan, The general solution of two-dimensional matrix Toda chain equations with fixed ends, \href{https://link.springer.com/article/10.1007/BF00750841}{Lett. in Math. Phys. \textbf{35}, 345, (1995)}.

\bibitem{2}

A. Davey and K. Stewartson, On three-dimensional packets of surface waves, \href{https://doi.org/10.1098/rspa.1974.0076}{Proc. Roy. Soc. A \textbf{338}, 101
(1974)}.


\bibitem{2a}

M. Boiti, J. P. Leon and F. Pempinelli, Multidimensional solitons and their spectral transforms, \href{https://doi.org/10.1063/1.529013
}{J. Math. Phys. \textbf{31}, 2612,
(1990)}.



\bibitem{3}

A. N. Leznov, Backlund transformation for integrable systems,
\href{https://doi.org/10.1007/BF01371391}{J. Russ. Laser Res.  \textbf{13}, 278 (1992)}.

\bibitem{3_1}
A. N. Leznov, Institute for High Energy Physics, Protvino preprint, IHEP-92-112 DTP, (1992).

\bibitem{5}

A. N. Leznov, A. B. Shabat and R. I.Yamilov, Canonical transformations generated by shifts in nonlinear lattices, \href{https://doi.org/10.1016/0375-9601(93)90197-8}{Phys. Lett. A \textbf{174}, 397, (1993)}.


\bibitem{6}

A. N. Leznov, On the complete integrability of a nonlinear system of partial differential equations in two-dimensional space, \href{https://doi.org/10.1007/BF01018624}{ Theor. Math. Phys. \textbf{42}, 343 (1980)}.






\bibitem{7}

A. N. Leznov and M. V. Saveliev, Progress in Physics
Vol. 15., p. 290 (Birkha\"user, Basel 1982).




\bibitem{4}

D. B. Fairlie and A. N. Leznov,
Durham University preprint DTP 33-93, (1993).

\bibitem{4_1}

A. N. Leznov, The new look on the theory of integrable systems, \href{https://doi.org/10.1016/0167-2789(95)00152-T}{Physica D \textbf{87}, 48 (1995)}.



\bibitem{quasi}
I. M. Gelfand, V. S. Retakh, Determinants of matrices over noncommutative rings,  \href{https://doi.org/10.1007/BF01079588}{Funct. Anal. Appl. \textbf{25}, 91 (1991)}.

\bibitem{quasi_1}

I. M. Gelfand, V. S. Retakh, A theory of noncommutative determinants and characteristic functions of graphs,
\href{https://doi.org/10.1007/BF01075044} {Funct. Anal. Appl. \textbf{26}, 1 (1992)}.

 


\end{thebibliography}
\end{document}